\documentclass{tran-l}
 \newtheorem{thm}{Theorem}

 \newtheorem{prop}[thm]{Proposition}
 \theoremstyle{definition}
 \newtheorem{defn}[thm]{Definition}
 \theoremstyle{remark}
 
 \numberwithin{equation}{subsection}

\begin{document}
\title{Entangled Graphs}

\author{Hadi Rahiminia, Massoud Amini}

\address{Department of Mathematics, Tarbiat Modarres University, P.O.Box
14115-175, Tehran, Iran} \email{rahiminia@modares.ac.ir,
mamini@modares.ac.ir}

\thanks{The authors would like to thank Professor Braunstein and his graduate students for
their kind help during the preparation of the original version of
this paper.}



\date{}

\dedicatory{}

\commby{}

\begin{abstract}
In this paper we prove a separability criterion for mixed states
in $\mathbb C^p\otimes\mathbb C^q$. We also show that the density
matrix of a graph with only one entangled edge is entangled.
\end{abstract}

\maketitle

\section*{Introduction}
One of the major problems in Quantum Mechanics is to characterize
entangled states of a quantum system. There are several partial
criteria for entanglement of mixed states [6], [8], [10], but
there is not yet a general criterion. Entanglement is connected to
the important concept of non-locality in Quantum Mechanics.
Entangled states are also useful in quantum cryptography and other
quantum information processing tasks [1],[9]. A mixed quantum state
is separable (and entangled, otherwise) if it can be written as a
convex combination of pure separable states . Solving the quantum
separability problem simply means determining whether a given
quantum state is entangled or separable.

Following [2], a class of states that are represented by the
density matrices of graphs are considered. It is shown in [2] that
certain classes of graphs always represent entangled (separable)
states. Also they have shown that a number of considered states
have an exactly fractional value of their concurrence — a measure
of entanglement of formation in small quantum systems.

A graph $G = (V,E)$ is a pair of a non-empty and finite set  $V$
(or $V (G)$) whose elements are called vertices; and a non-empty
set of unordered pairs of vertices $E$ (or $E(G)$), whose elements
are called edges. A loop is an edge of the form $\{v_i, v_i\}$,
for some vertex $v_i$. We assume that $E(G)$ does not contain only
loops [2]. Two distinct vertices $v_i$ and $v_j$ are adjacent if
$\{v_i, v_j\}\in E(G)$. The adjacency matrix of a graph on $n$
vertices $G$ is an $n \times n$ matrix $M(G)$, having rows and
columns labelled by the vertices of $G$, and $ij$-th entry defined
to be $1$ if $v_i$ and $v_j$ are adjacent; and $0$, otherwise. The
degree of a vertex $v_i$ is the number $d_G(v_i)$ of edges
adjacent to $v_i$. The degree of G is defined by $d_G
=\sum_{i=1}^n d_G(v_i)$. Note that $d_G = 2 |E(G)|$. The degree
matrix of $G$ is an $n\times n$ matrix $A(G)$, having $ij$-th
entry $d_G(v_i)$ if $i = j$; and $0$, otherwise. The Laplacian
matrix of a graph $G$ is the matrix $L(G)= A(G)- M(G)$. Note that
$L(G)$ does not change if one adds loops to or deletes loops from
$G$. The density matrix of a graph G is the matrix
$\sigma(G)=\frac{1}{d_G} L(G)$. A graph $G$ has $k$ components,
$G_1,G_2,\dots,G_k$,if there is an ordering of $V (G)$, such that
$M(G) = \sum_{i=1}^k M(G_i)$. In this case we write $G = G_1\oplus
G\oplus\dots\oplus G_k$. When no such decomposition exists except
for $k = 1$, $G$ is called connected. We refer the reader to [2]
for examples and more details.

Let $tr(A)$ be the trace of a matrix $A$. A density matrix $\rho$
is said to be pure if $tr(\rho^2) = 1$, and mixed, otherwise. [2,
Theorem 2.4] gives a necessary and sufficient condition on a graph
$G$ for $\sigma(G)$ to be pure.

If A is an $n\times n$ matrix, decomposed into $p^2$ blocks:
\[ A= \left[
         \begin{array}{rrrr}
              A^{1,1} & A^{1,2}&\ldots &A^{1,p}\\
              A^{2,1} & A^{2,2}&\ldots& A^{2,p}\\
              \vdots  \quad&  \vdots\quad&         &\vdots\quad\\
              A^{p,1} & A^{p,2}&\ldots& A^{p,p}\\
          \end{array} \right], \]
where each $A^{ij}$ is a $q\times q$ matrix and $n=pq$, then
(p,q)-partial transpose $A^{T_B}$ is given by:

\[A^{T_B}= \left[
         \begin{array}{rrrr}
              (A^{1,1})^T & (A^{1,2})^T&\ldots&(A^{1,p})^T\\
              (A^{2,1})^T & (A^{2,2})^T&\ldots& (A^{2,p})^T\\
               \vdots\quad &  \vdots\quad&         &\vdots\quad\\
              (A^{p,1})^T&(A^{p,2})^T&\ldots&(A^{p,p})^T
          \end{array} \right].\]

\smallskip

\section*{results}

Next two results positively answer two open problems raised in [2,
Conjecture 6.5]. A more general result in this direction is
obtained by Wu [11], but our method of proof is direct and gives a
better intuition in this special case.

\begin{thm} Let G be a graph $(\mid V\mid=pq)$. If G has only
one entangled edge, then $\sigma(G)$ is entangled.
\end{thm}

\begin{proof}
Let$P[\frac{1}{\sqrt{2}}(\mid ij\rangle-\mid
st\rangle)]$ be the only entangled edge of G such that $1\leq
i,s\leq p$ , $1\leq j,t\leq q$ and let G have all the possible
edges. We show that G is entangled. To prove this, we use the
separability s necessary condition (If $\sigma(G)$ is separable,
then $(\sigma(G))^{T_B}\geq 0$).

We look for a vector such that as X with $X^T\sigma (G)^{T_B}X<0$.
Consider the following
\[\quad\quad ij \quad\quad\quad\quad\quad\quad\quad st\]
\[x=[\frac{1}{2},\frac{1}{2},...,\frac{1}{2},\frac{p+q-1}{2(p+q)}
,\frac{1}{2},...,\frac{1}{2},\frac{p+q-1}{2(p+q)},\frac{1}{2},...,\frac{1}{2}]^T,\]

Now if we compute $X^T\sigma (G)^{T_B}X$ then after
simplification and using the fact that all the edges which are
not connected to vertices $x_{ij}$ and $x_{st}$ in the sum
$X^T\sigma (G)^{T_B}X$ arise to the terms in the form
$x^2+y^2-2xy (x=y=1/2)$, which is equal to zero, and the sum of
the other terms in $X^T\sigma (G)^{T_B}X$ for those edges that
are connected with $x_{ij}$ or $x_{st}$ may be written as:

\[(p+q-1)(x^2_{ij}+x^2_{st})+\sum_{k=1,k\neq j}^{q}x^2_{ik}+\sum_{k=1,k\neq i}^{p}x^2_{kj}+
\sum_{k=1,k\neq t,j}^{q}x^2_{sk}+\sum_{k=1,k\neq s,i}^{p}x^2_{kt}\]
\[-2x_{ij}(\sum_{k=1,k\neq j}^{q}x_{ik}+\sum_{k=1,k\neq i}^{p}x_{kj})-2x_{st}
(\sum_{k=1,k\neq t}^{q}x_{sk}+\sum_{k=1,k\neq s}^{p}x_{kt}).\qquad\qquad\qquad\]

Now since $x_{ij}=x_{st}=\frac{p+q-1}{2(p+q)}$ and the remaining
$x_{mn}$s $((m,n)\neq (i,j),(s,t))$ are $\frac{1}{2}$ after
substituting we have:

\[2(p+q-1)(\frac{p+q-1}{2(p+q)})^2+2(p+q-3)\frac{1}{4}-4(p+q-2)
(\frac{p+q-1}{4(p+q)})=-\frac{p+q-1}{2(p+q)^2},\] so
$\sigma(G)^{T_B}$ is not positive semi-definite, and therefore G
is entangled.

Next we suppose that there is a separable edge such as
$P[\frac{1}{\sqrt{2}}(\mid kl\rangle-\mid mn\rangle)]$ that is not
contained in G. If one of vertices of mentioned edge is $x_{ij}$ or
$x_{st}$, then, in the sum $X^T\sigma (G)^{T_B}X$, the term
$x^2_{mn}+x^2_{kl}-2x_{mn}x_{kl}$ appears that after substituting,
we get

\[(\frac{p+q-1}{2(p+q)})^2+\frac{1}{4}-\frac{p+q-1}{2(p+q)}=\frac{1}{4(p+q)^2}, \]

Now this expression is positive even if the edge is not in $G$,
and the proof goes as before. If the edge is not involving the
vertices $x_{ij}$ or $x_{st}$, then

\[x^2_{mn}+x^2_{kl}-2x_{mn}x_{kl}=\frac{1}{4}+\frac{1}{4}-\frac{1}{2},\]
and again we are done. To complete the proof we need to prove our
claim also for the following simple cases:

1. If the graph is just one edge that is entangled, trivially it is
entangled.

2. If all of separable edges of the graph are not connected with
vertices $x_{ij}$ and $x_{st}$, then for the vector X above, in the
sum $X^T\sigma (G)^{T_B}X$, only the expression
$x^2_{ij}+x^2_{st}-2x_{it}x_{sj}$ remains that after substitution,
it becomes

\[2(\frac{p+q-1}{2(p+q)})^2-2(\frac{p+q-1}{2(p+q)})(\frac{1}{2})=-\frac{p+q-1}{2(p+q)^2}<0.\]

Therefore all of the possible cases are considered, and we are done.
\end{proof}

\begin{thm}
If all the entangled edges of graph G are incident to the same
vertex, then G is entangled.
\end{thm}

\begin{proof}
We use Theorem 1. Let G have all the possible separable edges and
the edge $P[\frac{1}{\sqrt{2}}(\mid ij\rangle-\mid st\rangle)]$ be
one of the entangled edges and the vertex $x_{ij}$, be the common
vertex of the entangled edges. We prove that $\sigma(G)^{T_B}$ is
not positive semi-definite.

We omit all the entangled edges of graph G except
$P[\frac{1}{\sqrt{2}}(\mid ij\rangle-\mid st\rangle)]$ and call
the resulting graph H. We consider the vertex X as in Theorem 1.

From the proof of Theorem 1, the sum $X^T\sigma (H)^{T_B}X$ is
negative. Now if another edge of G such as
$P[\frac{1}{\sqrt{2}}(\mid ij\rangle-\mid mn\rangle)]$ is added to
H, the expression
\[x^2_{ij}+x^2_{mn}-2x_{in}x_{mj},\]
appears in the sum $X^T\sigma (H)^{T_B}X$, which after
substitution gives

\[(\frac{p+q-1}{2(p+q)})^2+\frac{1}{4}-\frac{1}{2}=-\frac{2(p+q)+1}{4(p+q)^2}<0. \]

Since the above expression is negative, if all of the omitted
entangled edges of G are added to H, the sum $X^T\sigma
(H)^{T_B}X$ remains negative and therefore G is entangled. Similar
to the proof of Theorem 1, one can show that the hypothesis that G
contains all the possible separable edges could be removed, and we
are done.
\end{proof}

\begin{defn}
A matrix is line sum symmetric if the $i$-th column sum is equal
to the $i$-th row sum for each $i$.
\end{defn}

We may use our technique combined with results of [11] to give
simpler proofs of some of the results proved in [2] with a
different method. The next three results are of this kind. For the
rest of the paper, $p$ and $q$ denote two arbitrary natural
numbers.

\begin{thm}
The density matrix of the tensor product of two graphs on $p$ and
$q$ vertices is separable in $C^p\otimes C^q$.
\end{thm}

\begin{proof}
Let $G$ be a graph on $p$ vertices and $H$ be a graph on $q$
vertices, with density matrices $\sigma(G)$ and $\sigma(H)$
respectively. By Theorem 8 of [11], it is enough to prove that
matrices $A^{ij}$ of $\sigma(G\otimes H)$ are line sum symmetric.
Any matrix $A^{ii}$ of $\sigma(G\otimes H)$, is symmetric and so
is sum line symmetric. Clearly we need only to show that the
matrices $M^{ij}$ of $M(G \otimes H)$ are line sum symmetric, for
$i\neq j$. Since $M(G\otimes H)=M(G)\otimes M(H)$, so each
$M^{ij}$ are equal to a multiplier of $M(H)$. Since $M(H)$ is
symmetric, we are done.
\end{proof}

Next we can decide on separability of the density matrix of two
special graphs, namely the complete graph $K_n$ on $n$ vertices,
and the star graph $K_{1,n-1}$ (see [2] for details).

\begin{prop}
$(i)$ For $n=pq$, the density matrix $\sigma(K_n)$ is separable in
$C^p\otimes C^q$.

$(ii)$ The density matrix of the star graph $K_{1,n-1}$ on
$n=pq\geq 4$ vertices is entangled in $C^p\otimes C^q$.
\end{prop}

\begin{proof}
$(i)$ Again note that for each of matrices $A^{ij}$ of
$\sigma(K_n)$, the $l$-th row sum is equal to the $l$-th column
sum, for $l=1,...,n$. Indeed, since the graph $K_n$ is complete,
all elements in $\sigma(K_n)$, expect for the diagonal elements,
are equal to -1, and we are one.

$(ii)$ Let $G=K_{1,n-1}$. It is obvious that in $\sigma(G)$, for
$i,j=2,...,p$, $A^{ij}$ is equal to $I_q$, if $i=j$, and $0$,
otherwise. By Theorem 3 of [11], it is enough to show that there
exists a row in $\sigma(G)^{T_B}$ with a nonzero row sum. Consider
$(p-1)q+1$-th row. But this row sum is clearly the summation of
the first row sums of matrices $(A^{pj})^T$, for $j=1,...,p$. This
last sum is now easily seen to be equal to $-q+1< 0$, therefore,
for $pq\geq4$, $G=K_{1,n-1}$ is entangled in $C^p\otimes C^q$.
\end{proof}

\begin{defn}
An e-matching is a matching having all edges entangled [2]. Each
vertex of an e-matching on $n=pq$ vertex can be labelled by an
ordered pairs $(i,j)$, where $1\leq i\leq p$ and $i\leq j\leq q$.
A pe-matching of a graph G is an e-matching spanning $V(G)$.
\end{defn}

\begin{thm}
Let $G$ be a graph on $n=2p$ vertices. If all the entangled edges
of $G$ belong to the same pe-matching, then $\sigma(G)$ is
separable in $C^2\otimes C^p$.
\end{thm}

\begin{proof}
Let $G$ be as above, we may divide $G$ into two graphs, consisting
of all separable edges and all entangled edges, respectively.
Let's call the second graph $H$. It is enough to show that $H$ is
separable.

The density matrix $\sigma(H)$ contains matrices
$A^{11}=A^{22}=I_q$, $A^{12}$ and $A^{21}$. Since the entangled
edges of $G$ form a pe-matching, so each row or column of matrices
$A^{12}$ and $(A^{21})$ has one $-1$ and all others zero. By
Theorem 7 of [11], $H$ is separable in $C^2\otimes C^p$, and we
are done.
\end{proof}

\end{document}